# Measuring the Gap Between Media Coverage and Public Information Demand: Evidence from the 2026 Lebanon Conflict

**Mohamed Soufan**
Independent Researcher & Data Scientist
Antalya, Turkey
ORCID: https://orcid.org/0009-0004-1705-5574
Email: m@soufan.tech
Website: https://www.soufan.tech

## Abstract

This study examines the relationship between media coverage and public information demand during the Lebanon conflict in March 2026. Using a dataset of 11,623 English-language news articles collected from the GDELT database and Google Trends data for searches conducted within Lebanon, the study compares the distribution of news coverage across topics with the distribution of public search interest. News headlines were filtered for relevance and classified into four categories: Conflict, Economy, Living Conditions, and Emigration. Public information demand was measured using Google Trends topic data for the same categories.

The results show a substantial divergence between news coverage and search interest. Conflict accounted for 94.9% of classified news coverage but only 36.9% of total search interest. In contrast, Economy, Living Conditions, and Emigration together accounted for 63.1% of search demand but only 5.1% of news coverage. Time series analysis indicates that search demand for economic and living conditions remained consistently elevated throughout the month rather than reacting to specific conflict events. These findings were robust to the exclusion of the peak conflict period (March 1–5), with Conflict coverage remaining at 94.9% and the information gap persisting across all three under-covered categories.

The findings suggest that during the study period, media coverage of Lebanon was heavily concentrated on military events, while public information demand was distributed across economic conditions, daily life, and emigration. This study contributes to agenda-setting research by providing a quantitative comparison between media agenda and public information demand during an active conflict period.

Keywords: agenda-setting, news coverage, Google Trends, information demand, Lebanon, conflict, news analysis

## Introduction

Armed conflicts generate intense media coverage, yet the topics emphasized by news organizations do not necessarily reflect the information needs of the public living through the conflict. During periods of crisis, people often seek information not only about military developments but also about economic conditions, daily life, and the possibility of leaving affected areas. Understanding whether news coverage aligns with public information demand is therefore important for understanding how information environments function during conflict.

This paper examines the relationship between news coverage and public information demand in the case of Lebanon during March 2026, a period characterized by frequent cross-border military activity and sustained media attention. The analysis compares the distribution of topics in English-language news coverage of Lebanon with the distribution of Google search interest from within Lebanon during the same period. By comparing these two information streams, the paper evaluates whether the media agenda and public information demand were aligned or divergent during an active conflict period.

The study is grounded in agenda-setting theory, which proposes that media coverage influences which issues the public perceives as important. This paper does not test the direction of influence. Instead, agenda-setting is used as a conceptual framework to examine whether the media agenda and public information demand were aligned during an active conflict.

Traditional agenda-setting research typically measures the relationship between media coverage and public opinion using surveys (McCombs & Shaw, 1972). More recent research has used digital trace data such as search queries and social media activity as proxies for public attention and information demand (Scharkow & Vogelgesang, 2011; Mellon, 2014). Recent computational research has also examined Lebanon's digital information environment using large-scale social media data, including work on how linguistic uncertainty shapes engagement in Arabic-language discourse (Soufan, 2026). Search data are particularly useful because they capture active information-seeking behavior rather than passive exposure to news content.

To compare media coverage and public information demand, news headlines and search topics were classified into four categories: Conflict, Economy, Living Conditions, and Emigration. These categories were selected to capture both conflict-related information and everyday concerns that may affect civilians during periods of instability. The analysis then calculates the distribution of news coverage and search interest across these categories and measures the gap between them.

The results show a large divergence between media coverage and public information demand during the study period. While news coverage was overwhelmingly focused on military events, search interest was distributed across economic conditions, living conditions, and emigration-related topics. These findings were consistent when the peak conflict period (March 1–5) was excluded from the analysis. This divergence suggests that the topics emphasized in English-language media coverage of Lebanon during the study period did not fully reflect the distribution of information demand observed in search behavior during the same period.

This paper contributes to the literature on agenda-setting and information behavior during conflict by providing a quantitative comparison between media coverage and public information demand using large-scale news and search datasets. Rather than examining whether media agenda influences public opinion, the study focuses on whether media coverage aligns with what people are actively seeking information about during a conflict period.

## Methodology

**Research Design**

This study adopts a comparative research design to examine the relationship between news coverage and public information demand during the Lebanon conflict in March 2026. It compares the distribution of topics in news coverage with the distribution of public search interest over the same period. The unit of analysis for media coverage is the individual news headline, whereas the unit of analysis for public information demand is Google Trends topic interest. The geographic focus is Lebanon, and the the analyzed period spans March 1 to March 31, 2026.

**News Data Collection**

News data were collected using the GDELT Document API, which provides access to global news articles (Leetaru & Schrodt, 2013).

The query term "Lebanon" was used to retrieve English-language articles published between March 1 and March 31, 2026. Because the GDELT API limits results to 250 articles per query, data were collected initially in one-day windows. When a query reached the API cap, the window was automatically split into 12-hour intervals and then into 6-hour intervals until all available articles were retrieved.

Duplicate articles were removed using two criteria: duplicate URLs and duplicate normalized titles. Titles were normalized by converting all text to lowercase and removing punctuation. After deduplication, the raw dataset contained 11,623 unique articles.

**Relevance Filtering**

A two-stage relevance filter was applied to remove articles that mentioned Lebanon only incidentally.

In the first stage, articles were required to contain at least one Lebanon-related anchor term, including *Lebanon*, *Lebanese*, *Beirut*, *Hezbollah*, *UNIFIL*, *South Lebanon*, *Tyre*, *Sidon*, or *Baalbek*.

In the second stage, articles primarily concerning other countries or conflicts were removed unless a strong Lebanon anchor was also present. Articles containing terms such as *Iran*, *Gulf*, *Yemen*, *Saudi*, *Canada*, *Europe*, *NATO*, *Ukraine*, or *Russia* were excluded unless the article clearly focused on Lebanon.

After relevance filtering, 2,428 articles remained in the dataset.

**Source Classification**

Each article was classified by source origin on the basis of its domain name. Sources were grouped into six categories: Lebanese, Israeli, International, Gulf/UAE, United States, and Other. This classification was used to describe the composition of the news dataset but was not used in the main gap calculations.

**Google Trends Data**

Public information demand was measured using Google Trends data. Google Trends data were collected using an Apify scraper for searches conducted within Lebanon between March 1 and March 31, 2026.

Google Trends topics were used rather than search terms because topics aggregate searches across multiple languages and related queries. This is especially important in Lebanon, where search activity occurs in Arabic, English, and French.

Four Google Trends topics were queried together in a single request so that values were normalized relative to one another on the Google Trends 0–100 scale.

**Table 1. Google Trends topics used for each category**

| Category | Google Trends Topic |
| --- | --- |
| Conflict | Hezbollah (Political Party) |
| Economy | Lira |
| Living Conditions | Rent |
| Emigration | Passport |

These topics were selected to represent the main non-conflict information needs related to economic conditions, living costs, and emigration intentions.

A fifth category related to safety and refugees was tested but showed near-zero search volume in Lebanon and was therefore excluded from the analysis.

**News Classification**

News headlines were classified into four categories: Conflict, Economy, Living Conditions, and Emigration. Classification followed a priority rule in which the cause of an event took precedence over its consequences. For example, a headline describing an airstrike that caused displacement was classified as Conflict rather than Living Conditions. Categories were evaluated in a fixed order: Conflict first, followed by Economy, Emigration, and Living Conditions. Each headline was assigned to the first category whose criteria it satisfied.

All 2,428 headlines were classified using GPT-4o via the ChatGPT interface. The model was provided with the full category definitions, the fixed priority order, and explicit disambiguation rules in a structured prompt. To validate the classifications, a stratified random sample of 100 headlines was independently labeled by two human annotators. One annotator was the lead researcher, and the second was an independent annotator with no prior knowledge of the study hypotheses. The human labels matched the GPT-4o classifications for all 100 headlines, yielding 100% agreement. Articles that did not fit any category were labeled as Other and excluded from percentage calculations. In total, 534 articles (22% of the dataset) were classified as Other.

**Gap Calculation**

The analysis compares the distribution of news coverage and search interest across the four categories.

For each category, news percentage was calculated as:

News% = (Number of articles in category / Total classified articles) × 100

Search percentage was calculated by dividing each topic's monthly average Google Trends value by the sum of the monthly averages across all four topics.

The information gap was then calculated as:

Gap = Search% − News%

A positive gap indicates that public information demand exceeded media coverage for a given topic, whereas a negative gap indicates that media coverage exceeded public information demand.

**Reproducibility / Code Availability**

All data collection and analysis scripts used in this study are publicly available at: https://github.com/mohamedsoufan/lebanon-media-search-gap.

# Results

**Dataset Overview**

The initial GDELT query returned 11,623 unique English-language news articles containing the term "Lebanon" during March 1–31, 2026. After the two-stage relevance filter was applied, 2,428 articles remained. All retained headlines were then classified into five outcomes: Conflict, Economy, Emigration, Living Conditions, and Other. Of these, 1,894 articles were assigned to one of the four substantive categories used in the analysis, whereas 534 articles (22.0%) were classified as Other and excluded from percentage calculations. Search demand was measured using Google Trends topic data for four categories over the same 31-day period.

**Distribution of News Coverage**

News coverage was overwhelmingly concentrated on conflict-related events. Of the 1,894 classified articles, 1,798 were labeled as Conflict, representing 94.9% of the analyzed news corpus. Living Conditions accounted for 66 articles (3.5%), Emigration for 23 articles (1.2%), and Economy for only 7 articles (0.4%).

Figure 1 shows the number of news articles per category in the classified news corpus.

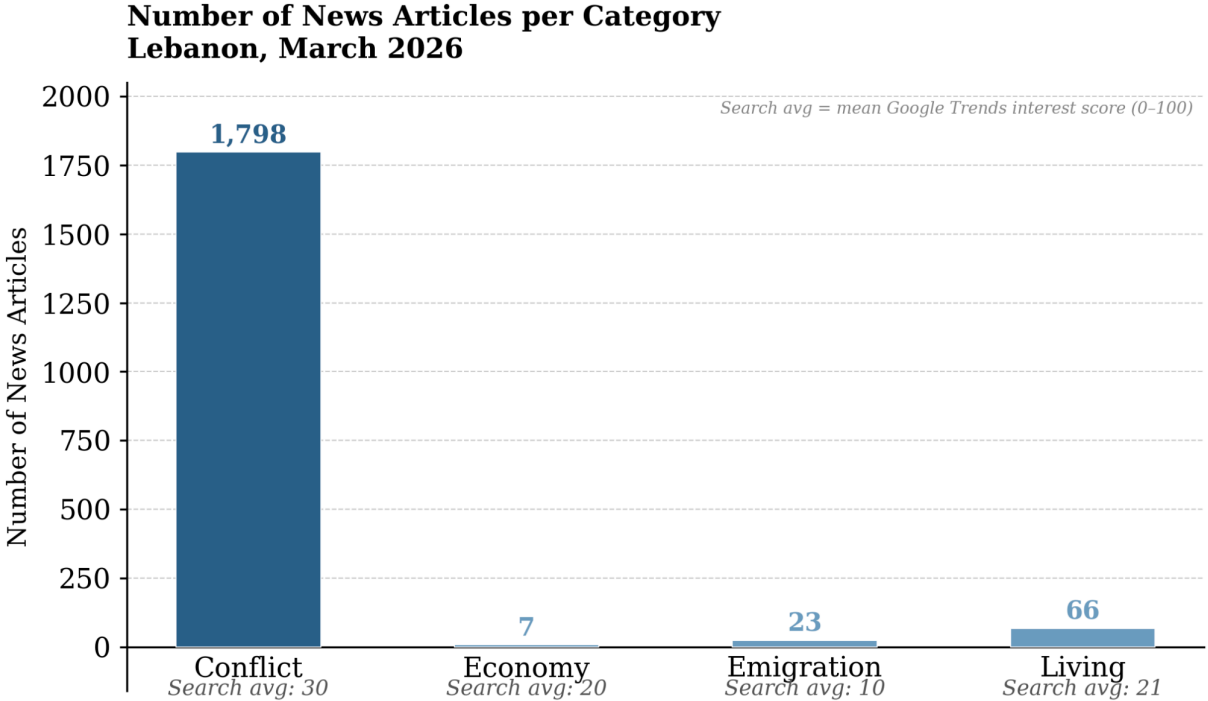

This distribution indicates that English-language media coverage of Lebanon during the observation window was dominated almost entirely by military developments. Non-conflict concerns received minimal attention by comparison. The very small number of Economy articles should be interpreted cautiously, as some economically relevant coverage may have been absorbed into Conflict or Other under the study's classification rules. If so, the measured economy gap is likely conservative rather than overstated.

**Composition of News Sources**

The filtered news corpus consisted primarily of sources classified as Other (54.5%), followed by International outlets (22.2%), Israeli outlets (10.3%), Lebanese outlets (6.9%), United States outlets (4.3%), and Gulf/UAE outlets (1.7%). This source composition is relevant for interpretation because Israeli and many international sources tended to cover Lebanon primarily through a military lens, which may have increased the relative share of conflict coverage in the English-language corpus.

**Distribution of Public Information Demand**

Google search interest from within Lebanon was distributed much more evenly across topics than the news agenda. Conflict, measured through the Hezbollah topic, had a monthly average interest score of 30.3 and accounted for 36.9% of total search share. Living Conditions, measured through the Rent topic, had an average score of 21.1 and a 25.8% share. Economy, measured through the Lira topic, had an average score of 20.2 and a 24.6% share. Emigration, measured through the Passport topic, had an average score of 10.5 and a 12.7% share.

Figure 2 compares the distribution of news coverage and public search demand across the four categories.

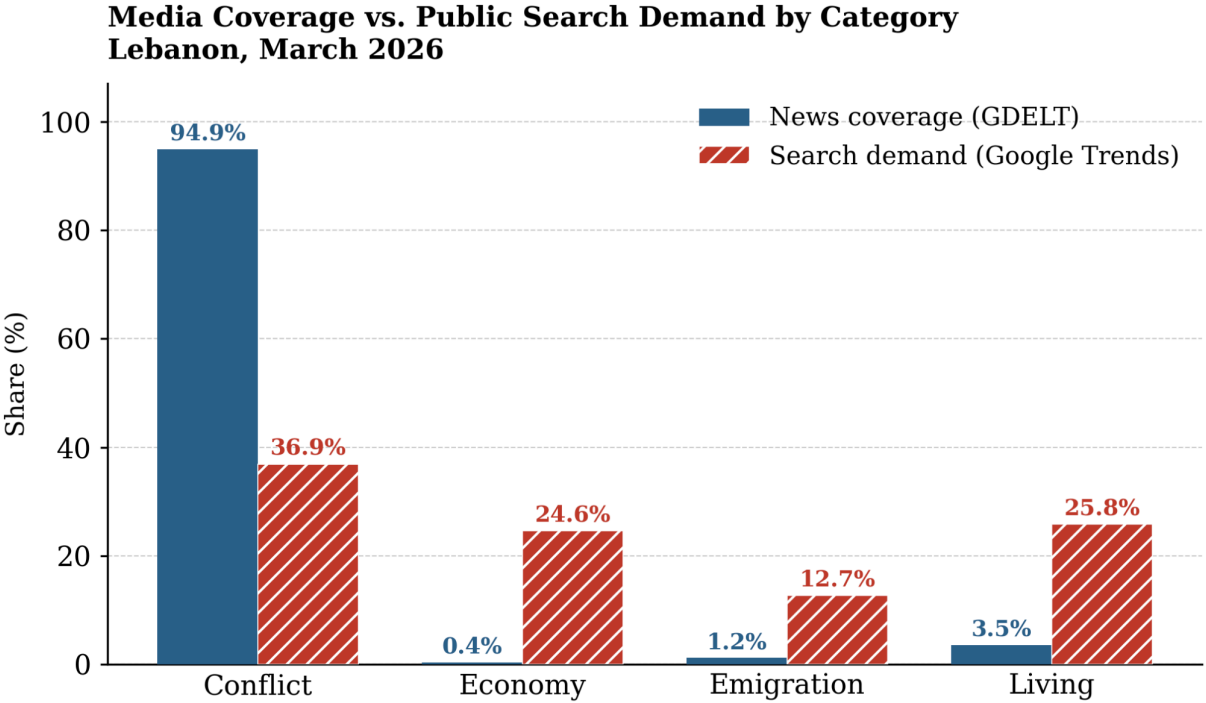

In contrast to the news dataset, no single topic dominated search activity. Although Conflict generated the largest share of search interest, Economy, Living Conditions, and Emigration together accounted for most total search demand.

**Media–Search Gap**

A substantial divergence emerged between the distribution of news coverage and the distribution of search interest. Conflict was the only category over-represented in news coverage relative to search demand. It accounted for 94.9% of news coverage but only 36.9% of search share, yielding a gap of −58.0 percentage points.

The coverage gap between media coverage and search demand for each category is shown in Figure 3.

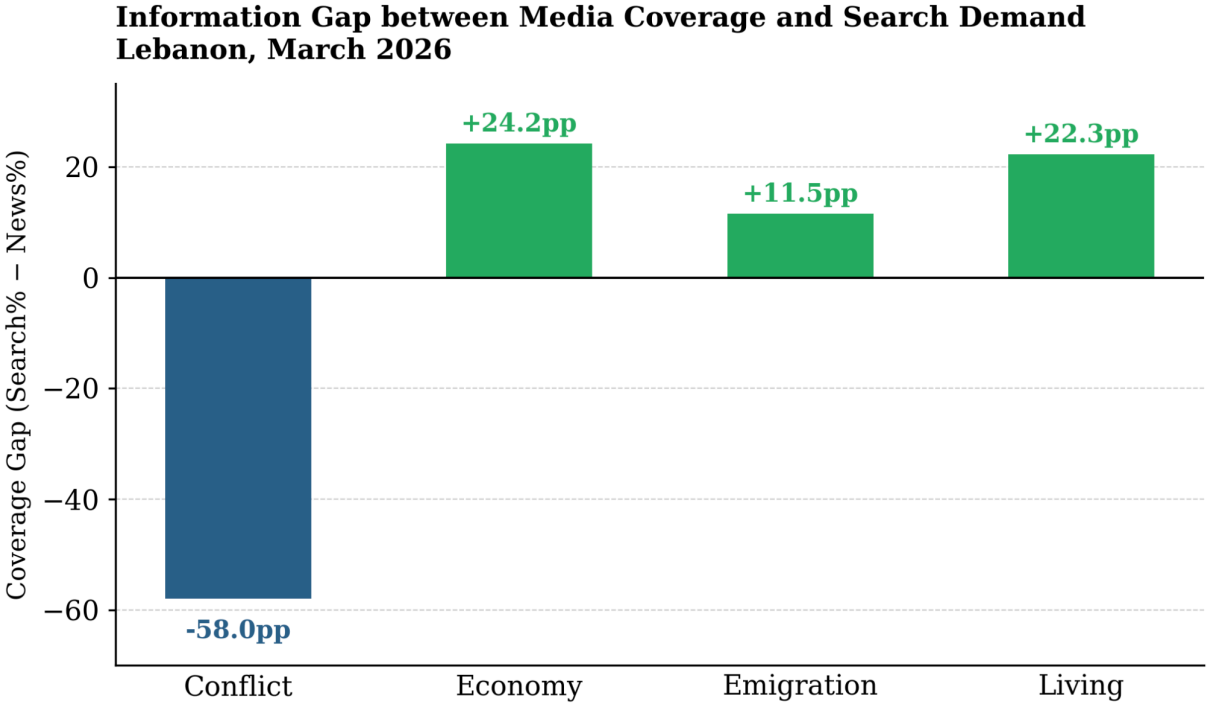

All three non-conflict categories were under-represented in news coverage. Economy exhibited the largest positive gap: it accounted for only 0.4% of news coverage but 24.6% of search share, yielding a gap of +24.2 percentage points. Living Conditions accounted for 3.5% of news coverage and 25.8% of search share, producing a gap of +22.3 percentage points. Emigration accounted for 1.2% of news coverage and 12.7% of search share, producing a gap of +11.5 percentage points.

Taken together, Economy, Living Conditions, and Emigration represented 63.1% of total search demand but only 5.1% of total news coverage. This contrast constitutes the central empirical finding of the study.

**Time Series Patterns**

Temporal patterns also differed between the media agenda and search demand. Conflict-related news coverage peaked during March 1–5, which was also the period with the highest daily article counts in the news corpus. By contrast, non-conflict search demand did not appear to be driven solely by major conflict spikes.

Figure 4 shows daily conflict news coverage and public search demand over the 31-day period.

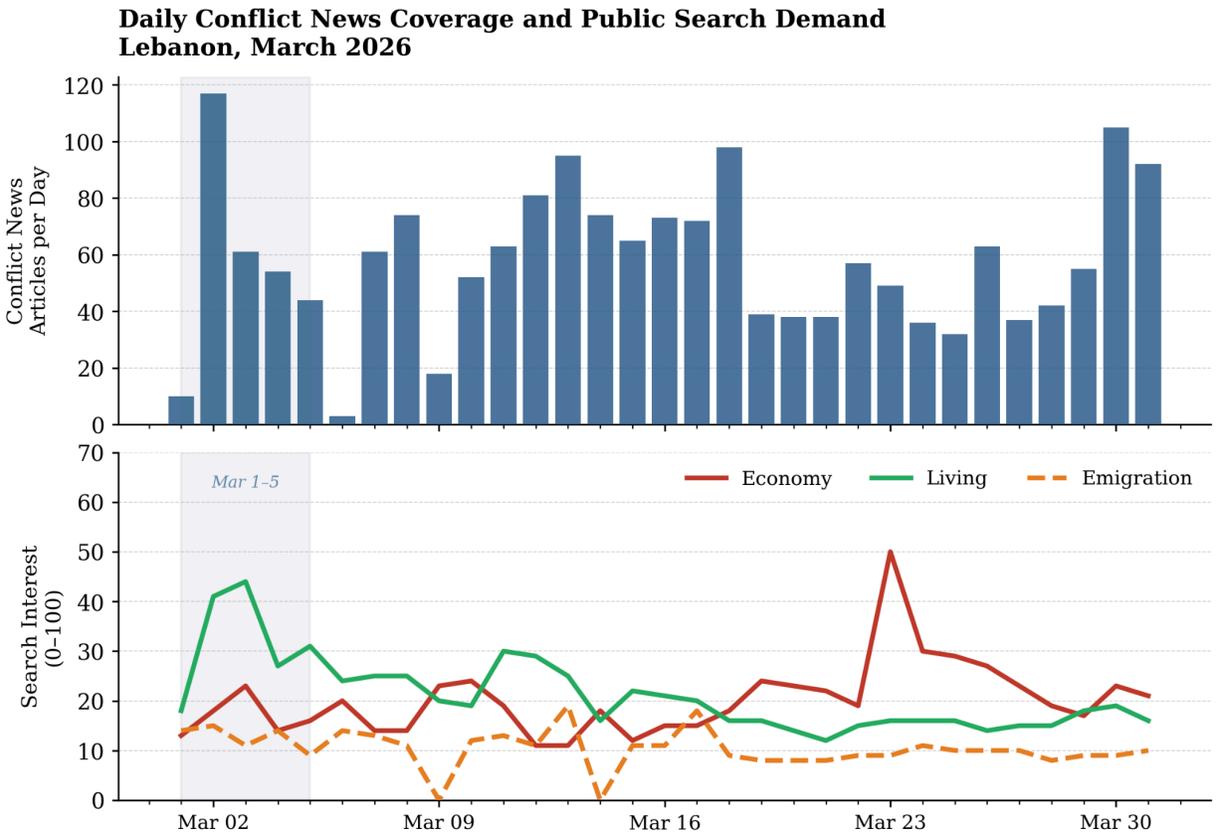

Economy-related search interest remained elevated throughout the month and showed a notable spike around March 23. Living Conditions search interest was highest at the beginning of the month and then gradually declined. Emigration-related search interest remained relatively stable at a lower level throughout March. These patterns suggest that search demand for non-conflict topics was persistent over time rather than simply reactive to major military events.

No clear co-movement was observed between spikes in conflict news coverage and economy- or emigration-related search interest. Instead, the search series for these topics suggest ongoing background information needs that were sustained across the month.

**Validation**

To evaluate the reliability of the headline classification procedure, a stratified random sample of 100 headlines was independently labeled by two human annotators, one of whom was the lead researcher and the other an independent annotator with no prior knowledge of the study hypotheses. Annotators completed their labels independently before results were compared. The human labels matched the GPT-4o classifications for all 100 headlines. This produced 100% agreement between model and human annotations, as well as 100% inter-annotator agreement within the validation sample.

The high level of agreement likely reflects the relatively explicit structure of the category definitions, the fixed priority order, and the fact that many headlines were topically unambiguous.

**Robustness Check**

To assess whether the findings were driven primarily by the most intense conflict days, all articles published during March 1–5 were excluded and the analysis was repeated on the reduced dataset. The main result remained unchanged. Conflict coverage remained 94.9% of the classified corpus, and the conflict gap widened slightly from −58.0 to −60.4 percentage points. Economy, Living Conditions, and Emigration all remained under-covered by substantial margins.

This robustness check indicates that the observed divergence between media coverage and information-seeking behavior was not simply an artifact of the first week of escalation. The imbalance persisted even after the highest-volume conflict period was removed.

## Discussion

The findings reveal a clear divergence between the distribution of English-language media coverage of Lebanon and the distribution of Google search interest from within Lebanon during the the 31-day period. Although conflict dominated the news agenda almost completely, search demand was distributed across conflict, economy, living conditions, and emigration. The central pattern is therefore not that conflict was unimportant in search behavior. Rather, the imbalance lies in the scale of emphasis: conflict accounted for nearly all news coverage, whereas non-conflict concerns accounted for most measured search demand.

Within an agenda-setting framework, this pattern suggests that the media agenda and search interest were not closely aligned during the study period. The English-language news corpus overwhelmingly emphasized military events, whereas search behavior indicated sustained interest in practical and civilian-oriented concerns. This does not establish that media coverage failed in a normative sense, nor does it imply that search behavior represents the complete public agenda. It does, however, indicate that the issue priorities visible in media output differed substantially from those visible in active information-seeking behavior.

One implication is that conflict reporting may capture only part of the informational environment experienced during war. Military developments are highly newsworthy, especially for international and cross-border outlets, but they are not the only concerns that attract attention during crisis periods. The elevated search shares for the lira, rent, and passport topics suggest that economic stability, daily living conditions, and the possibility of leaving the country remained salient throughout the month. Passport-related search activity may reflect both emigration intent and routine administrative needs, but its sustained elevation during an active conflict period suggests that departure-related concerns were, at minimum, one component of search behavior.

The economy category result is particularly notable. Economy accounted for only seven classified articles, yet the lira topic represented nearly one quarter of total search demand. This was the largest positive gap observed in the study. Because some economically relevant headlines may have been absorbed into Conflict or Other under the classification rules, this estimate should be interpreted conservatively. Even so, the direction of the pattern is unambiguous: economic concerns were far more prominent in search behavior than in the news corpus analyzed here.

The time series patterns reinforce this interpretation. Conflict coverage was concentrated most heavily in the first days of March, but economy-related search interest remained elevated throughout the month, and living-condition searches were also substantial, especially early in the period. Emigration-related search demand remained lower than the other non-conflict categories, but still far exceeded its share of coverage. These patterns suggest that non-conflict information needs were not limited to isolated reactions to battlefield events. Instead, they appeared as persistent background concerns during the conflict period.

The robustness check further strengthens this interpretation. Excluding March 1–5 did not materially alter the findings. Conflict coverage remained at 94.9% of the classified corpus, and all three non-conflict categories remained substantially under-covered relative to search demand. This indicates that the observed gap was not simply a by-product of a short burst of intense escalation at the beginning of the month. Rather, the divergence persisted across the broader 31-day period.

These findings should also be interpreted in light of the composition of the dataset. The news corpus consisted of English-language coverage, with substantial shares from international and Israeli sources and a relatively small share from Lebanese outlets. This matters because English-language coverage of Lebanon is likely shaped by external news values, especially conflict intensity, diplomacy, and regional security. Lebanese domestic Arabic-language media may present a different issue balance, potentially giving greater attention to economy and daily life. The present findings therefore apply most directly to the English-language media agenda captured through GDELT, rather than to all media coverage of Lebanon.

The analysis also highlights the value of search data as a measure of search interest. Unlike surveys, search traces capture active information-seeking behavior in real time. They do not reveal motivations or user identity, and they remain imperfect proxies, but they provide a useful way to observe which kinds of information attract sustained attention during crisis periods. In the present study, search data make visible a dimension of public concern that is largely absent from the news distribution alone. More broadly, the study demonstrates that combining large-scale news archives with search data offers a tractable method for measuring information gaps during active conflicts, where survey-based approaches are often difficult to implement in real time.

At the same time, the comparison should not be over-interpreted. News percentages and search percentages are different kinds of measures: one is based on article counts, whereas the other is based on normalized search intensity. The proxy topics are also incomplete representations of broader categories. Hezbollah does not capture all conflict-related searches, and lira, rent, and passport do not exhaust the domains of economy, living conditions, and emigration. For this reason, the results should be understood as evidence of directional imbalance rather than as precise estimates of the full gap between media attention and public concern.

Taken together, the findings suggest that during March 2026, the English-language media agenda on Lebanon was narrower than the search demand visible through search behavior. The dominant news focus on conflict did not eliminate interest in other issues; instead, it coexisted with substantial and persistent demand for information about everyday survival, economic strain, and exit options. This divergence constitutes the main contribution of the study and underscores the importance of examining conflict information environments beyond battlefield coverage alone.

## Limitations

This study has several limitations that should be considered when interpreting the findings.

First, the Google Trends data used in the analysis measure relative search interest rather than absolute search volume. The values are normalized on a 0–100 scale within the selected query set, meaning they indicate the distribution of attention across the four chosen topics rather than the total volume of searches conducted within Lebanon. Accordingly, the search measures should be interpreted as proxies for relative search interest rather than direct counts of information-seeking activity.

Second, the comparison in this study relies on two different types of measures. News coverage is represented by article counts, whereas Google Trends reflects normalized search intensity. These measures are not directly equivalent. The analysis therefore does not assume that the two series are directly comparable in scale or magnitude; instead, they are used as parallel indicators to compare the relative prominence of issue categories in media output and search behavior.

Third, the Google Trends categories are measured through proxy topics rather than exhaustive conceptual categories. Hezbollah was used as a proxy for conflict-related search interest, lira for economy, rent for living conditions, and passport for emigration. These topics were selected because they showed the clearest and most consistent signal in Lebanon during March 2026, but they do not capture all searches related to each domain. For example, not all economic concern is expressed through searches about the lira, and passport-related searches may reflect both emigration-related concerns and routine administrative needs. The results should therefore be interpreted as indicative of directional patterns rather than complete measurements of each category.

Fourth, the news corpus includes only English-language coverage retrieved through GDELT. This means the analysis captures the English-language media agenda on Lebanon rather than the full Lebanese domestic media environment. Arabic-language Lebanese outlets may give greater attention to economic conditions, service disruptions, or daily civilian concerns than the English-language corpus analyzed here. The findings should therefore not be generalized to all media coverage of Lebanon.

Fifth, the GDELT API imposes a 250-result limit per query window. Although the collection procedure mitigated this constraint by automatically splitting high-volume periods into smaller time windows, peak news days may still have been undercounted. This concern is most relevant for the first week of March, when conflict coverage was especially intense. However, the robustness check excluding March 1–5 produced substantively similar results, which reduces concern that the main findings were driven solely by this limitation.

Sixth, the source composition of the corpus may have amplified conflict coverage. Israeli outlets accounted for 10.3% of the filtered news dataset, and such outlets tend to cover Lebanon primarily through a military lens. International English-language outlets may show a similar emphasis. Because the analysis did not re-estimate the gap after removing these sources, it is possible that the observed conflict share is somewhat inflated relative to a corpus with a larger domestic Lebanese component. Replicating the analysis with a corpus limited to Lebanese and neutral international sources would be a useful robustness check for future work.

Seventh, 22.0% of the filtered articles were classified as Other and excluded from the percentage calculations. These articles were typically diplomatic background pieces, international political coverage that mentioned Lebanon only partially, cultural items, or opinion content that did not fit the four substantive categories. Excluding them improved category clarity, but it also means that the final gap analysis is based on a reduced subset of the filtered corpus rather than all relevant Lebanon-related articles.

Eighth, the extremely small number of Economy articles in the classified news set suggests that this category may be sensitive to classification boundaries. Some headlines related to reconstruction, infrastructure damage, or financially consequential war effects may have been classified as Conflict because the coding rules prioritized cause over consequence. This means the reported economy gap may be conservative. The direction of the finding is clear, but the exact magnitude should be interpreted with caution.

Ninth, the validation exercise produced unusually high agreement: the two human annotators and GPT-4o agreed on all 100 sampled headlines. Annotators completed their labels independently before results were compared. While this likely reflects the clarity of the category definitions, the fixed priority order, and the relative directness of many headlines, the validation sample remains limited. High agreement in this context supports the reliability of the coding scheme, but it does not eliminate the possibility of borderline cases elsewhere in the full corpus.

Finally, the study is limited to a single month, March 2026. This was an active conflict period, and the issue balance observed here may not represent quieter periods in Lebanon or other conflict settings. The findings should therefore be interpreted as period-specific rather than universal. The study identifies a strong media–search divergence during March 2026, but further work across multiple months and conflict intensities would be needed to determine how stable this pattern is over time.

## Conclusions

This study compared the topic composition of English-language news coverage with the pattern of Google search interest from within Lebanon. Across 1,894 classified news articles and four Google Trends topic categories, the results showed a clear divergence between the media agenda and public information demand during an active conflict period. Conflict dominated the news corpus, accounting for 94.9% of classified coverage, while Economy, Living Conditions, and Emigration together accounted for only 5.1%. In contrast, these three categories represented 63.1% of total search demand, while Conflict accounted for 36.9%. These findings were robust to the exclusion of the peak conflict period (March 1–5), with Conflict coverage remaining at 94.9% and the information gap persisting across all three under-covered categories.

The main contribution of the study is empirical. The results indicate that during March 2026, the issue priorities visible in English-language media coverage of Lebanon did not closely match the distribution of information demand visible in search behavior. While military developments overwhelmingly structured the news agenda, search demand remained distributed across both conflict-related and civilian concerns, particularly economic conditions and daily life. This pattern remained stable even after excluding the peak conflict period at the start of the month.

The study also makes a methodological contribution by demonstrating that large-scale news archives and search data can be combined to examine media–public information gaps during active crises. This approach offers a practical alternative in settings where survey-based measures of public attention may be difficult to collect in real time.

At the same time, the findings should be interpreted within the study's scope. The analysis is limited to one month, one country context, an English-language news corpus, and a small set of proxy search topics. Accordingly, the results should be understood as evidence of a strong divergence during a specific conflict period rather than as a universal claim about conflict coverage more broadly.

Future research could extend this design across longer time periods, compare English-language and Arabic-language media agendas, test alternative proxy topics, and examine whether similar media–search gaps appear in other conflict settings. Additional robustness checks that vary source composition could also clarify how much the observed imbalance depends on the structure of the news corpus itself.

Overall, the study shows that during March 2026, the English-language media agenda on Lebanon was far narrower than the pattern of information-seeking behavior visible in search data. Conflict dominated coverage, but it did not dominate information demand to the same extent. The gap identified here suggests that understanding crisis information environments requires attention not only to what the media emphasize, but also to what people actively seek to know.

## Supplementary Materials

The headline classification definitions, category rules, priority order, and GPT-based classification prompt used in the study are provided as Supplementary Materials.

## Data Availability Statement

Due to database access limitations and source licensing restrictions, the raw news article dataset cannot be redistributed. The list of article URLs, classification outputs, and derived aggregate datasets used in the analysis are available from the corresponding author upon reasonable request. The data collection and analysis code used in this study are publicly available in the project GitHub repository.

## Author Contributions

Mohamed Soufan conducted all aspects of the study, including conceptualization, methodology, data collection, software development, formal analysis, investigation, data curation, visualization, and writing. Manual annotation assistance used for validation is acknowledged below.

## Acknowledgments

The author thanks Aya Soufan for her assistance with manual annotation during the validation of the headline classification.

## Funding

This research received no external funding.

## Institutional Review Board Statement

Ethical review and approval were waived for this study because the research analyzed publicly available news articles and aggregated search trend data and did not involve interaction with human participants or access to private or sensitive personal data.

## Informed Consent Statement

Not applicable.

## Conflicts of Interest

The author declares no conflict of interest.